\documentclass[twocolumn,nature]{revtex4-2}

\usepackage[utf8]{inputenc}
\usepackage{epsf,graphicx}
\usepackage{natbib}
\usepackage{multirow}
\usepackage{amsfonts,amsmath,gensymb,amssymb}
\usepackage{siunitx}
\usepackage[bottom]{footmisc}
\usepackage{lipsum}
\usepackage{dcolumn}
\usepackage{xcolor}
\usepackage{upgreek}
\usepackage{float,ulem}

\usepackage{bm}
\usepackage{hyperref}
\hypersetup{
colorlinks=true, 
citecolor=magenta,
breaklinks=true, 
urlcolor= blue, 
linkcolor= blue, 
bookmarksopen=true, 
}
\usepackage{natbib}
\usepackage{color}
\usepackage[type={CC},modifier={by-nc-sa},version={4.0},]{doclicense}
\definecolor{linkcolor}{rgb}{0,0,0.6}

\begin{document}

\title{Effects of nonlinearity on Anderson localization of surface gravity waves}	

\author{Guillaume Ricard}
\affiliation{Universit\'e Paris Cité, CNRS, MSC, UMR 7057, F-75013 Paris, France}

\author{Filip Novkoski}
\affiliation{Universit\'e Paris Cité, CNRS, MSC, UMR 7057, F-75013 Paris, France}

\author{Eric Falcon}
\affiliation{Universit\'e Paris Cité, CNRS, MSC, UMR 7057, F-75013 Paris, France}

\begin{abstract}
Anderson localization is a multiple-scattering phenomenon of linear waves propagating within a disordered medium. Discovered in the late 50s for electrons, it has since been observed experimentally with cold atoms and with classical waves (optics, microwaves, and acoustics), but whether wave localization is enhanced or weakened for nonlinear waves is a long-standing debate. Here, we show that the nonlinearity strengthens the localization of surface-gravity waves propagating in a canal with a random bottom. We also show experimentally how the localization length depends on the nonlinearity, which has never been reported previously with any type of wave. To do so, we use a full space-and-time-resolved wavefield measurement as well as numerical simulations. The effects of the disorder level and the system's finite size on localization are also reported. We also highlight the first experimental evidence of the macroscopic analog of Bloch's dispersion relation of linear hydrodynamic surface waves over periodic bathymetry. 
\end{abstract}

\maketitle


\noindent \textbf{Introduction}\\
Wave propagation in nonhomogeneous media has led to astonishing phenomena for over a century, especially in the field of solid-state physics~\cite{Bragg1913,Bloch1929,Anderson1958}. For example, in a periodic medium, wave scattering at each change in the spatial property of the medium can result in a forbidden frequency band for which waves cannot propagate (Bragg reflection)~\cite{Bragg1913}. In a random medium, interferences between multiply-scattered waves by random defects can halt wave propagation and trap or localize it within the medium. This linear phenomenon, known as Anderson localization, was first predicted for electrons in condensed matter~\cite{Anderson1958,Mott1967}, and has since been observed in various domains~\cite{Sheng1990,Lagendijk2009} ranging from cold atoms~\cite{Billy2008,Roati2008,KondovScience2011,JendrzejewskiNatPhys2012} to classical waves such as optics~\cite{WiersmaNature1997,StorzerPRL06,PertschPRL04,SchwartzNature2007,Lahini2008}, microwaves~\cite{DalichaouchNature91,ChabanovNature00}, acoustics~\cite{Hodges1982,Depollier1986,Desideri1993,Hu2008}, as well as for random time-varying media~\cite{Sharabi2021,Apffel2022}. For surface waves on a liquid, only a few laboratory studies carried out in the 80s, have demonstrated Anderson localization of linear surface waves over random bathymetry~\cite{Belzons1987,Belzons1988} and the existence of a forbidden frequency band in the periodic case~\cite{Heathershaw1982,HaraJFM1987,Guazzelli1992}. More recent experiments report water wave attenuation in canals involving resonators~\cite{LorenzoJFM2023,EuvePRL2023} or corrugated sidewalls~\cite{Zorkani1991,ZhangAIPAdv18,Wang2022}, but do not focus on wave localization. It is the same for experiments on water waves over a two-dimensional periodic structure showing Bloch-like patterns~\cite{Torres99,Torres2003}, superlensing and self-collimation effects~\cite{Hu04,Shen05}.

Experimental studies of Anderson localization mainly concern linear waves, although there is a long-standing debate about whether nonlinearity enhances or weakens wave localization by disorder. Some theoretical studies predict that Anderson localization persists for continuous nonlinear waves~\cite{Frohlich1986,Albanese1988}, but can be destroyed for nonlinear pulses~\cite{Li1988,Kivshar1990,PikovskyPRL08,IvanchenkoPRL11}, while experimental studies with nonlinear waves are scarce and limited to optics~\cite{PertschPRL04,SchwartzNature2007,Lahini2008}, acoustics~\cite{Mckenna1992}, or superfluid Helium~\cite{Hopkins1996,Hopkins1998}. In the context of surface gravity waves, the effects of nonlinearity on Anderson localization are still unknown both experimentally and theoretically, despite many potential applications regarding coastal protection from large-amplitude waves~\cite{Bailard1992,Zhang2012,LorenzoJFM2023}.  

Here, we show that nonlinearity enhances the localization phenomenon of surface gravity waves propagating over a random bathymetry in a canal. This result is achieved thanks to a full space-and-time-resolved wavefield measurement, that goes significantly beyond the indirect measurements used forty years ago to show localization of linear water waves~\cite{Belzons1987,Belzons1988}. 
Using this highly resolved technique available today, we also experimentally observe the Bloch dispersion relation of linear waves over a periodic potential, which has not been reported so far for water waves. We also use numerical simulations to corroborate our experimental observations. Moreover, we report theoretically a smooth transition between Bragg scattering (periodic bottom) to Anderson localization (random bottom) when the disorder level is increased. Finite-size effects of the system on the localization length are also studied.\\

\begin{figure*}[t!]
    \centering
       \includegraphics[width=1\linewidth]{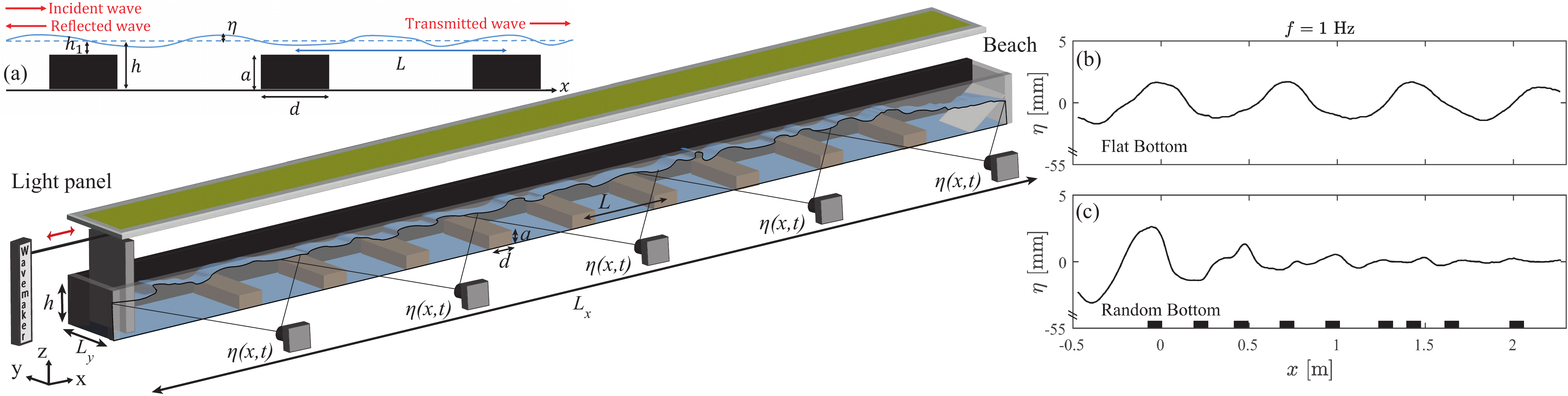}
    \caption{\textbf{Experimental setup to study Anderson localization of surface gravity waves in a 4-m canal over a random or periodic bathymetry.} $a=4$ cm, $d=8$ cm, $L=25$ cm, and $h=5.5$ cm. Insets: \textbf{(a)} Schematics for a periodic bottom. \textbf{(b,c)} Typical experimental wave amplitude, $\eta(x)$, along the first 2.8 m of the canal for \textbf{(b)} a flat bottom and \textbf{(c)} a random bottom with $L=25+\delta$~cm. Note that vertical axes are cut. Forcing frequency $f=1$~Hz and wave steepness $\epsilon=0.015$. Black rectangles indicate the bar locations in the canal.}
    \label{setup}
\end{figure*} 

\noindent \textbf{Results}\\
\noindent \textsf{Experimental setup}\\
The experimental setup is illustrated in Fig.~\ref{setup} (see photograph in the Supp.~Sect.~1 and Video~1). It consists of a transparent canal of length $L_x=4$~m and of width $L_y=18$~cm. The canal is filled with water to a depth $h=5.5$~cm. At one end, a linear motor drives a paddle to generate monochromatic surface waves in a frequency range $f\in[0.42,1.80]$~Hz (corresponding to wavelengths $\lambda \in[0.36,1.70]$~m). These waves propagate along the $x$ axis and are dissipated by an 80-cm long beach, made of 2 cm stones, at the other end. This beach ensures that almost no reflection occurs at the canal end. The spatiotemporal measurement of the surface elevation $\eta(x,t)$ is made by the use of five cameras (Basler, 20 fps, 6 Mpx) regularly spaced laterally along the canal, each filming an 86~cm side view with an angle of 4$^\circ$ following the method in ref.~\cite{Redor2020}. The surface is illuminated from the top to generate a strong contrast between the water surface and a black lateral background. The horizontal and vertical resolutions are 0.28~mm/pixel. The five cameras are synchronized in time to film simultaneously the water surface along the canal. A 7-cm overlap is made between each camera view to ensure an accurate reconstruction of the whole wavefield. The nonlinearity is quantified by the wave steepness as $\epsilon=\left\langle\sqrt{\int |\partial\eta/\partial x|^2dx/L_x}\right\rangle_t$. We choose $\epsilon \in[0.01, 0.04]$ for the flat bottom to avoid wavebreaking. The typical rms amplitude of waves is 2 mm. To create a spatial-dependent bathymetry, $N=9$ rectangular aluminum bars, of width $d=8$~cm and of height $a=4$~cm, are placed along the bottom of the canal. For a periodic bathymetry, the bars are placed regularly every $L=25$~cm and thus $L/ \lambda \in[0.2, 0.7]$. Hence, the water depth changes accordingly: $h=5.5$~cm between the bars ($kh \in[0.20, 0.96]$) and $h_1=h-a=1.5$~cm over each bar ($kh \in[0.06, 0.26]$). The random bathymetry is created from the periodic lattice by moving each bar by a length $\delta$ chosen randomly and uniformly in the range $\delta \in [-\kappa L/2, \kappa L/2]$ for each bar. $\kappa$ thus quantifies the level of disorder from 0 (periodic case) to 1 (fully random). Note that changing the distance $L+\delta$ between successive bars instead of other parameters (e.g., bar heights or widths) is the easiest and most efficient way of achieving the effect of randomness~\cite{Belzons1988}. The values of $L$, $N$, and $\kappa=0$ or 1 are fixed experimentally and are varied for theoretical predictions. The complete randomness ($\kappa=1$) will be referred to as the random bottom afterward. Along with the experiments, numerical simulations are performed using the fully-nonlinear Boussinesq equations solver, as part of Basilisk software~\cite{Popinet2015} (see Methods).

\vspace{0.2cm}
\noindent \textsf{Linear theoretical dispersion relation}\\
Because of the change in bathymetry, the dispersion relationship of linear gravity waves reads
\begin{align}
    \omega^2=\left\{ 
        \begin{array}{ll}
            gk_1\tanh(k_1h_1) \ &\mathrm{over\  the\ steps,} \\
            gk\tanh(kh) \ &\mathrm{elsewhere,}  
        \end{array}
         \right.
         \label{relat_disp_kh_kh1_eq}
\end{align}
with $\omega=2\pi f$ the angular frequency, $k$ and $k_1$ the wave numbers between and above the steps, respectively. Note that capillary effects are negligible since only long waves are considered here. Using the transfer matrix method over a periodic lattice~\cite{Huang2005,Ye2003} (see Supp.~Sect.~5) and the Bloch theorem, i.e., $\eta(x)=U(x)\exp{(iKx)}$ with $U(x)$ a periodic function of period $L$ and $K(\omega)$ the Bloch wave number~\cite{Bloch1929}, the dispersion relationship of linear waves over a periodic bottom reads~\cite{AN2004,Huang2005,Zhang2012,Ye2003}
\begin{equation}
    \begin{split}
    \cos(KL)=&\cos[k_1(\omega)d]\cos[k(\omega)(L-d)]\\
    &-\cosh(2\zeta)\sin{[k_1(\omega)d]}\sin{[k(\omega)(L-d)]},\\
    \end{split}
    \label{K_bloch_eq}
\end{equation}
with $\zeta=\ln{\left(\sqrt{k_1/k}\right)}$ and with $k_1(\omega)$ and $k(\omega)$ given by inverting Eq.~\eqref{relat_disp_kh_kh1_eq}. Equation~\eqref{K_bloch_eq} thus replaces the usual dispersion relation of surface linear waves over a flat bottom of Eq.~\eqref{relat_disp_kh_kh1_eq}, taking into account the reflections induced by each step. In the case of random bathymetry, no dispersion relationship is predicted, to the best of our knowledge, because of the invalidity of the Bloch theorem.

\begin{figure*}[t!]
    \centering
    \includegraphics[width=1\linewidth]{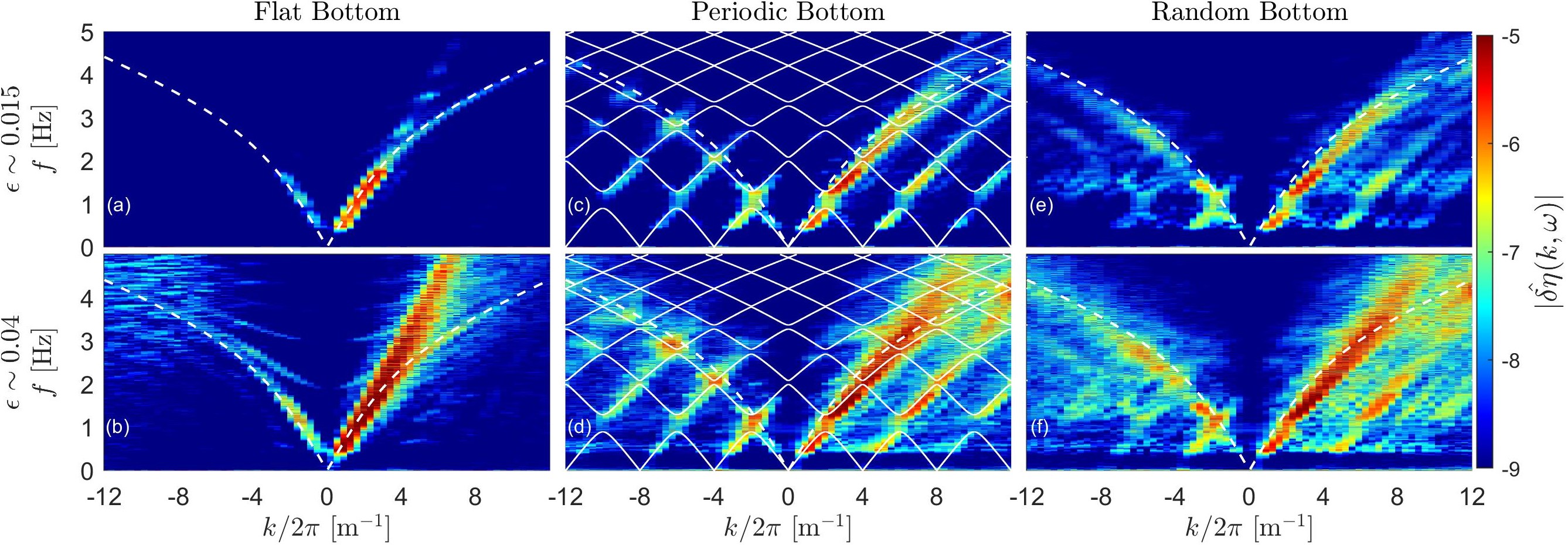}
    \caption{\textbf{Full spatiotemporal spectrum of the surface wavefield, $|\hat{\delta\eta}(k,\omega)|$.} Weak forcing ($\epsilon=0.015$): \textbf{(a)} Flat bottom, \textbf{(c)} periodic bottom, and \textbf{(e)} random bottom. Strong forcing ($\epsilon=0.040$): \textbf{(b)} Flat bottom, \textbf{(d)} periodic bottom, and \textbf{(f)} random bottom. Sweep-sine forcing $f\in[0.42, 1.8]$~Hz. Dashed-white lines: Eq.~(\ref{relat_disp_kh_kh1_eq}b). Solid-white lines: Eq.~\eqref{K_bloch_eq} with a $1/L$ spatial periodicity. Logarithmic-scale colorbar.}
    \label{Spatio_temp}
\end{figure*}

\vspace{0.2cm}
\noindent \textsf{Spatiotemporal wave spectrum}\\
Examples of typical wavefields along the canal are reported in Fig.~\ref{setup}b for a flat bottom and in Fig.~\ref{setup}c for a random bottom, both for the same forcing frequency and wave steepness. Waves are halted for the random case after typically 1 m whereas they propagate over the flat bottom. This wave localization by the random bathymetry occurs within a frequency band ($1.1\pm0.2$~Hz) (see Supp.~Sect.~2 and Video~2). To experimentally measure the wave dispersion relation, we use a sweep-sine forcing (from 0.42~Hz to 1.8~Hz  during 5~min) rather than monochromatic forcing, which gives similar results but requires significantly more individual measurements. The wavemaker amplitude $A(t)$ is chosen to keep a constant wavemaker acceleration, $A(t)f(t)^2$, over time $t$ ensuring a constant nonlinearity $\epsilon$ for each probed frequency $f$. Increasing nonlinearity, $\epsilon$, is made by increasing the wavemaker acceleration (i.e., increasing $A$ using a constant $f$) as $A(t)f(t)^2\in[4,30]$~mm/s$^2$. The evolution of the surface elevation $\eta(x,t)$ along the canal is shown in Supp.~Sect.~2, and Video~2 and Video~3, for small and large nonlinearity, respectively. From the signal $\eta(x,t)$, we then compute the spatiotemporal spectrum of the surface elevation $|\hat{\eta}(k,\omega)|$ using a double Fourier transform. As it provides similar but noisier results, we rather plot the spectrum $|\hat{\delta\eta}(k,\omega)|$ of the surface wavefield difference $\delta\eta(x,t)=\eta(x+dx)-\eta(x)$ with $dx$ the spatial resolution. The results are reported for a weak forcing $\epsilon=0.015$ in Fig.~\ref{Spatio_temp}a, \ref{Spatio_temp}c, and~\ref{Spatio_temp}e for a flat, a periodic, and a random bathymetry, respectively. The results for a strong forcing $\epsilon=0.040$ are reported in Fig.~\ref{Spatio_temp}b, \ref{Spatio_temp}d and~\ref{Spatio_temp}f. 

For the flat bottom and a weak forcing (Fig.~\ref{Spatio_temp}a), the wave energy follows as expected the linear gravity dispersion relation of Eq.~(\ref{relat_disp_kh_kh1_eq}b) (dashed lines). With almost no energy present for $k<0$ (less than 10$\%$ of the incident waves is detected for both linear and nonlinear cases), we conclude that the beach is efficient in preventing wave reflection. Increasing the forcing (Fig.~\ref{Spatio_temp}b) leads to strong nonlinear effects: the major part of the energy is now spread along a straight line $\omega\sim k$, characteristic of the presence of coherent structures such as trains of solitons. Note that less energetic sloshing branches emerge (from $k\simeq0$ and $f\ne0$).

In the case of a periodic bottom, the waves undergo reflections at each step and the classical dispersion relation of Eq.~\eqref{relat_disp_kh_kh1_eq} (dashed lines) is no longer valid. For a weak forcing (linear waves) (Fig.~\ref{Spatio_temp}c), the wave energy is found to be in good agreement with the Bloch dispersion relation $K(\omega)$ of Eq.~\eqref{K_bloch_eq} (solid lines). Reflections of the incident waves at each step lead to the emergence of stationary waves. Remarkably, almost no energy is found within a frequency gap ($f\in[0.9,1.3]$~Hz) as expected by the Bloch theory (energy is about one to two orders of magnitude smaller than outside the band gap). The agreement between the Bloch dispersion relation and experiments has been observed in optics~\cite{Engelen2005,Dubey2017} and is here shown for the first time for hydrodynamic surface waves using spatiotemporal measurements.  This result is also replicated by the numerical simulations thus validating them (see Supp.~Sect.~3). For a stronger forcing (Fig.~\ref{Spatio_temp}d), the Bloch dispersion relation, which is a linear prediction, is less respected due to nonlinear effects leading to more energy appearing within the band gap.

In the case of a random bottom (Fig.~\ref{Spatio_temp}e and~\ref{Spatio_temp}f), the loss of periodicity implies that the Bragg scattering and Bloch dispersion relation are theoretically no longer valid. This is indeed observed experimentally as the wave energy broadens more strongly, filling up more homogeneously the ($k$,$\omega$) plane (compare Fig.~\ref{Spatio_temp}d and~\ref{Spatio_temp}f). Note that the random changes of the bathymetry imply numerous reflections, generating a large amount of dispersive waves, instead of the standing waves found for the periodic case.

\vspace{0.2cm}
\noindent \textsf{Wave evolution along the canal}\\
We now focus on the variation of the wave amplitude along the canal to infer the localization length. To do so, we plot the maximum $\eta_{max}(x)$ of the oscillating wave amplitude at each location $x$ in the canal for different incident frequencies. In the case of a flat bottom (Fig.~\ref{setup}b and Supp.~Sect. 2), almost no decrease of the wave amplitude occurs as viscous effects are negligible. Indeed, using an exponential fit $\eta_{max}(x)\sim \exp{(-x/l_d)}$, we obtain a dissipative length $l_d\sim20\pm10$~m, in agreement with theory (see Supp.~Sect.~6), which is much longer than the canal length $L_x$. As dissipation has the same signature as localization [i.e., an exponential decrease of the wave amplitude, see Eq.~\eqref{expo_dec_loca_eq}], dissipation has to be negligible to be able to highlight the localization phenomenon. 

\begin{figure}[t!]
    \centering
                       \includegraphics[width=1\linewidth]{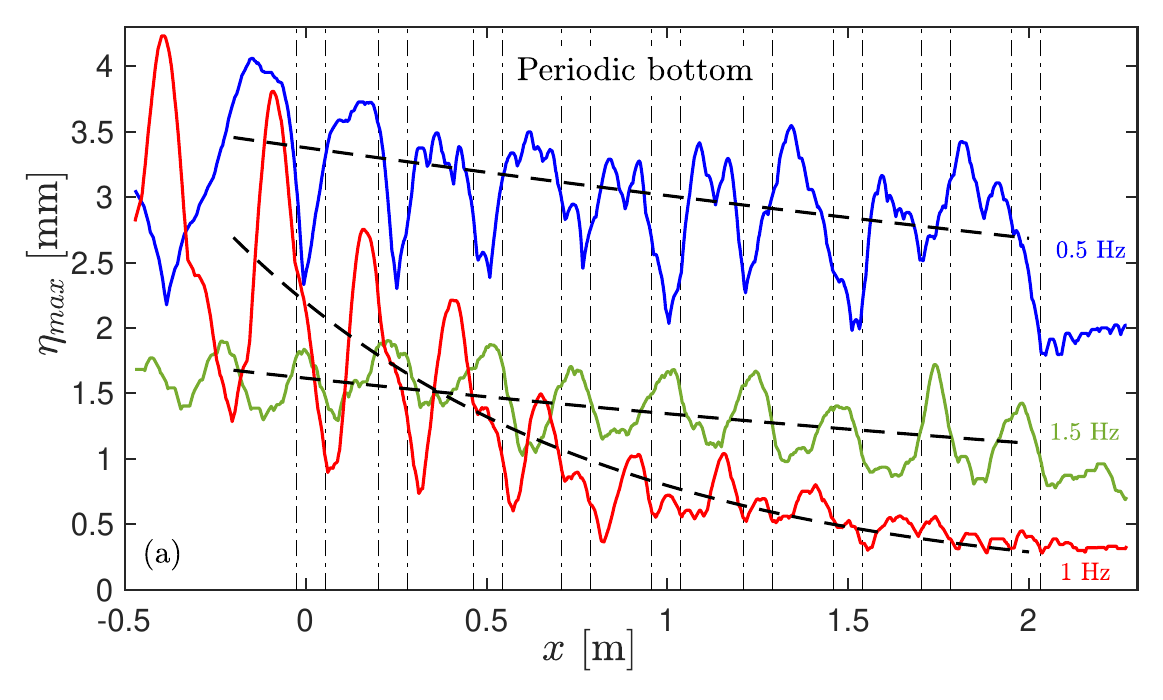}
       \includegraphics[width=1\linewidth]{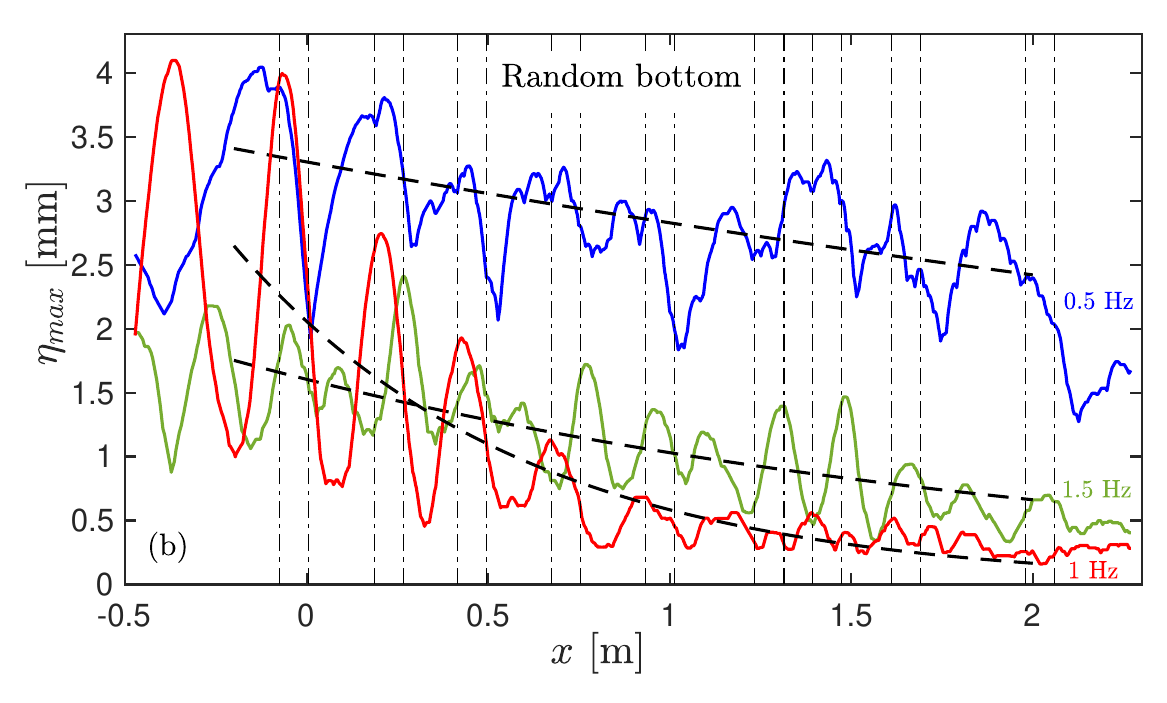}
    \caption{\textbf{Wave amplitude maxima $\mathbf{\eta_{max}}$ along the canal.} Different forcing frequencies $f=0.5$ (blue), 1 (red), and 1.5 (green)~Hz, steepness $\epsilon=0.015$. \textbf{(a)} Periodic bottom, \textbf{(b)} random bottom. Dashed lines: best exponential fits $\eta(x)=\eta_0\exp{(-x/\xi)}$. Vertical dash-dotted lines: location of the bars in the canal.}
    \label{eta_max}
\end{figure}

The variations of $\eta_{max}(x)$ are represented in Fig.~\ref{eta_max} for three different forcing frequencies $f=0.5$ (blue), 1 (red) and 1.5~Hz (green) in the periodic (a) and random (b) cases. $\eta_{max}$ decreases slightly along $x$ for $f=0.5$ and 1.5~Hz as these frequencies are located before and after the reminiscent band gap, respectively (see Fig.~\ref{Spatio_temp}c). Conversely, a strong decrease of $\eta_{max}(x)$ is found for $f=1$~Hz, a frequency inside the reminiscent band gap. In the periodic case [see Fig.~\ref{eta_max}a], for each frequency, we also observe that the bathymetry implies strong oscillations of $\eta_{max}(x)$ as a consequence of standing waves emerging from the reflections of the incident waves at each step. Note that the position of these standing wave nodes depends on the incident frequency. Indeed, for a frequency below the band gap (0.5~Hz), the maxima of $\eta_{max}$ is roughly located between two successive steps (schematized by the vertical-dashed lines), whereas it is located just above the steps for a frequency above the band gap (1.5~Hz). For an incident frequency within the band gap (1~Hz), the maxima of the standing waves occur just before the steps and their minima just after. To sum up, when the standing wavelength is in phase with the lattice spatial period, no attenuation occurs and the bathymetry does not influence the wave propagation, whereas a significant decrease of the wave amplitude occurs, due to intense reflections, when the standing wave is out-of-phase with the lattice. The random bathymetry leads to qualitatively similar results but with more randomness in the reflected waves generated [see Fig.~\ref{eta_max}b] and a more pronounced exponential decay (see below).

\begin{figure}[t!]
       \includegraphics[height=5cm,width=8.5cm]{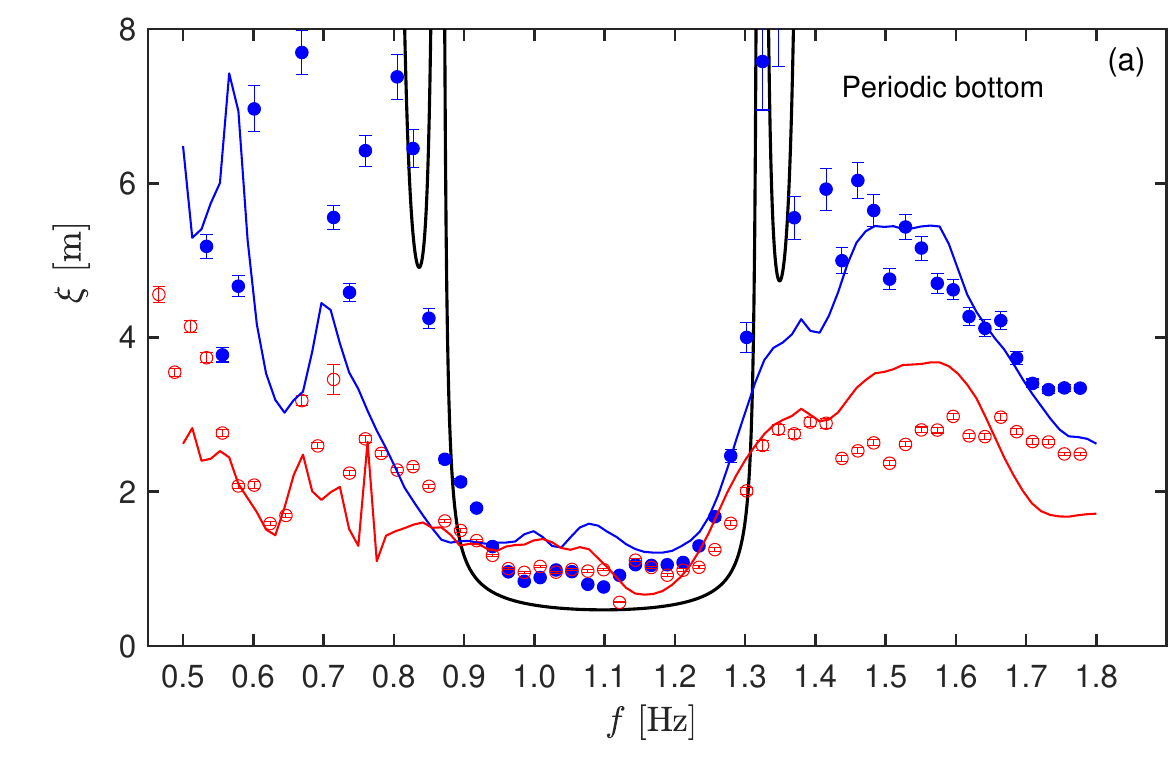}\\
    \includegraphics[height=5cm,width=8.8cm]{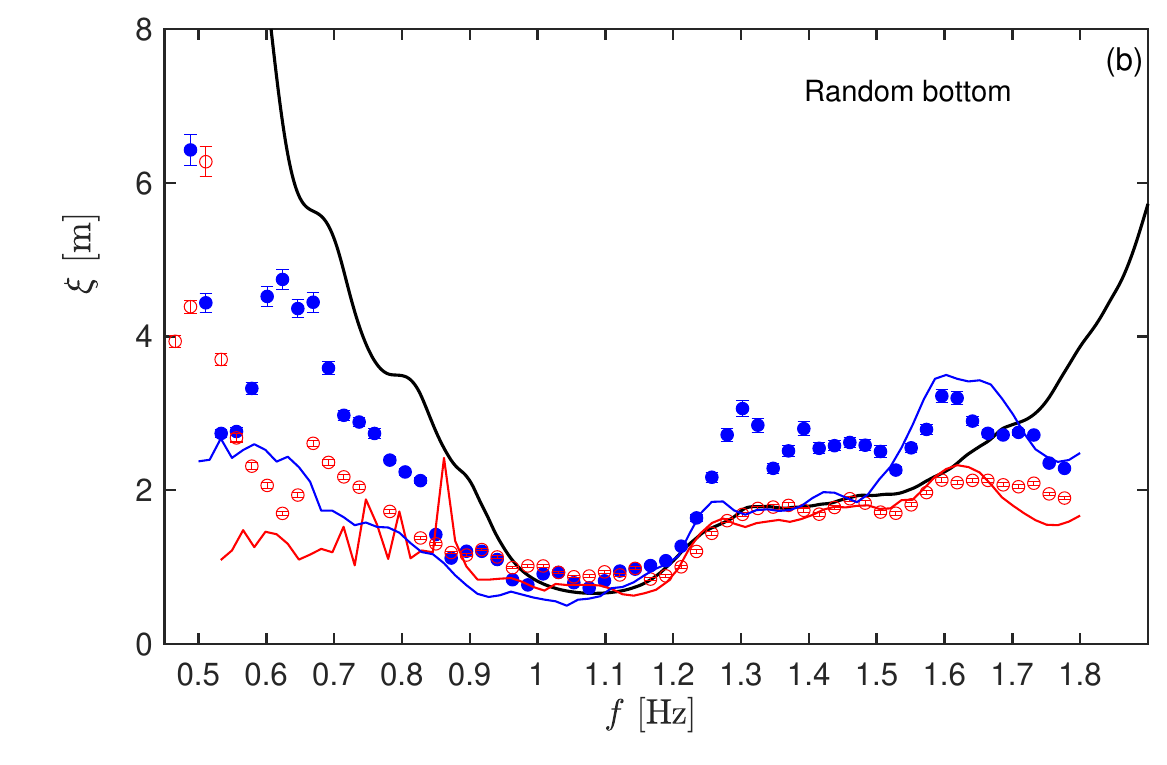}
        \begin{tabular}{ll}
    \includegraphics[width=0.5\linewidth]{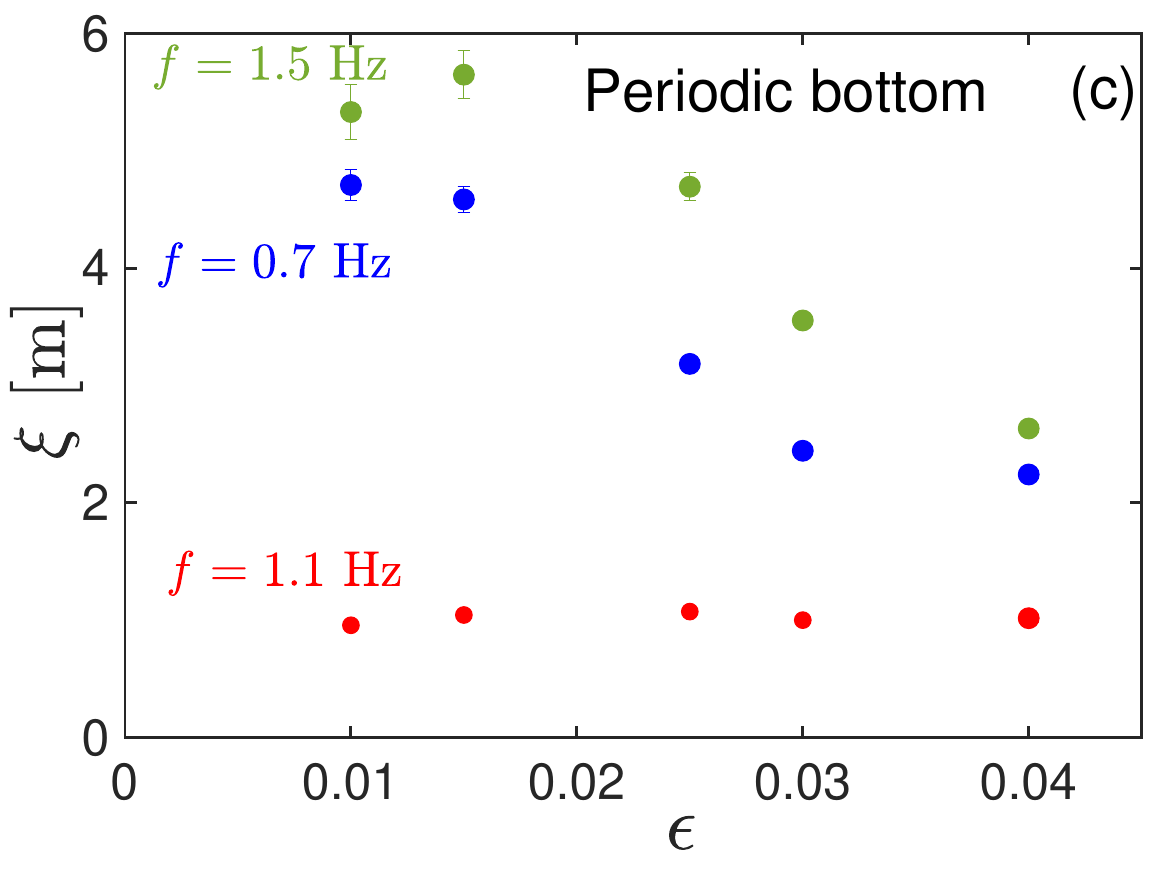}&
    \includegraphics[width=0.5\linewidth]{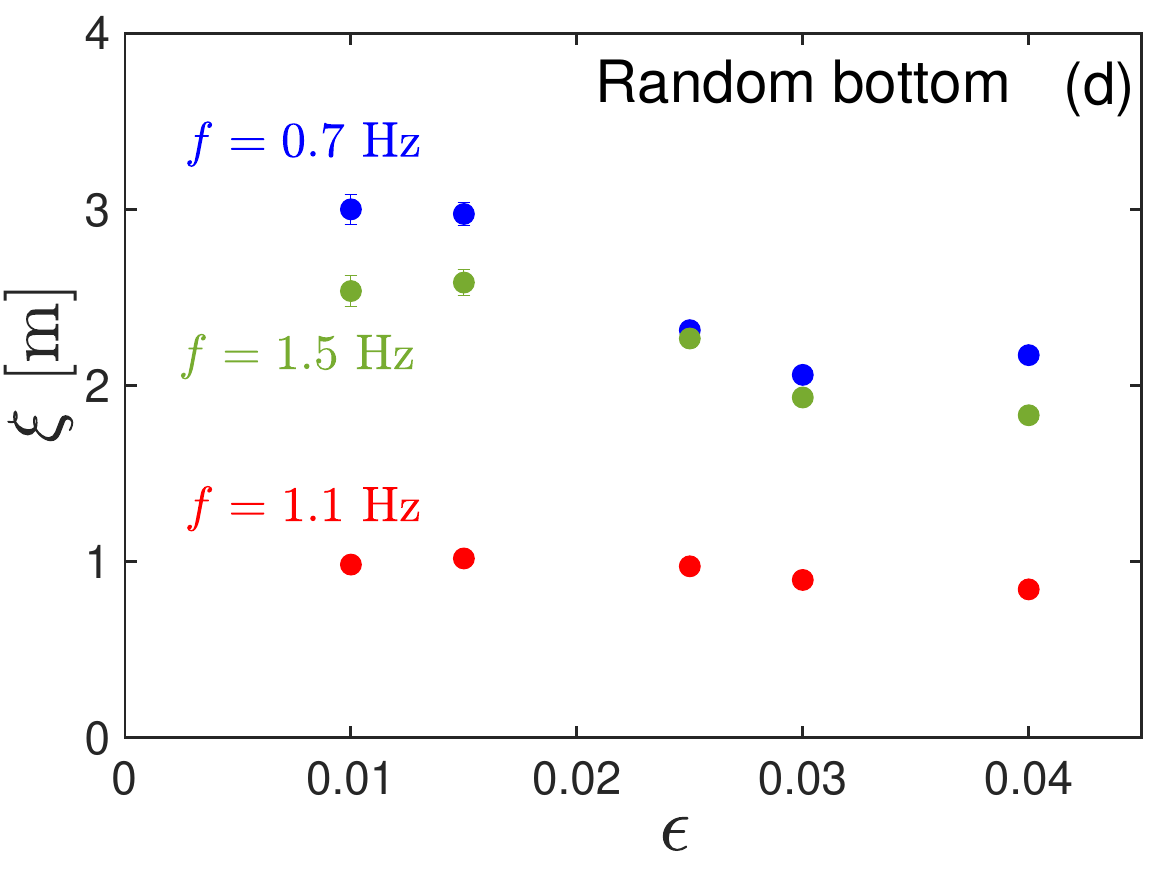}
    \end{tabular}
        \caption{\textbf{Localization length $\xi$ and nonlinearity $\epsilon$.} \textbf{(a)} $\xi$ versus frequency $f$ for  $\epsilon=0.015$ (blue bullets for experiments and blue lines for simulations) and $\epsilon=0.040$ (red empty symbols for experiments and red lines for simulations) for the periodic case. \textbf{(b)} Same for the random case. Solid-black lines: linear theoretical predictions of Eq.~\eqref{loca_th_eq}. \textbf{(c)} $\xi$ versus $\epsilon$ for a frequency lower (0.7~Hz), larger (1.5~Hz), and within (1.1~Hz) the theoretical band gap for the periodic case. \textbf{(d)} Same for the random case. Error bars are related to the exponential fit error (Fig.~\ref{eta_max}).}
    \label{xi}
\end{figure}

\vspace{0.6cm}
\noindent \textsf{Localization length and nonlinearity}\\
Despite the presence of standing waves, $\eta_{max}(x)$ can be fitted (dashed lines in Fig.~\ref{eta_max}) with an exponential spatial decay as
\begin{equation}
    \eta(x)=\eta_0\exp{(-x/\xi)}.
    \label{expo_dec_loca_eq}
\end{equation}
 The corresponding values of the localization length $\xi$ are then plotted for different values of the incident frequency $f$ in the periodic (Fig.~\ref{xi}a) and random cases (Fig.~\ref{xi}b), for weak, $\epsilon=0.015$ (blue bullets for experiment and blue lines for simulations), and strong forcing, $\epsilon=0.040$ (red empty symbols for experiments and red lines for simulations). We compare the data with the values of $\xi(f)$ predicted by a linear theory (solid black line) computed by the transfer matrix method (see Supp.~Sect.~5) as~\cite{Huang2005}
\begin{equation} 
    \xi (\omega)=\frac{NL}{\ln(\mathcal{T}_{11})},
    \label{loca_th_eq}
\end{equation}
with $N$ the number of steps, $NL$ the total lattice length, and $\mathcal{T}_{11}$ the inverse of the transmission coefficient between the wave amplitude before and the one after the lattice. The theoretical localization length $\xi$ thus depends on the incident wave frequency and the bathymetry features (see Supp.~Sect.~5). It turns out that some incident wave frequencies are not impacted by the bathymetry (full transmission), whereas others (close to the Bragg resonance~\cite{Bragg1913}, i.e., $\lambda \sim 2L=0.5$~m) can no longer propagate in the lattice (no transmission) and are localized.

For the weak forcing, we observe a good agreement between the experimental and theoretical results. For the periodic bathymetry and a weak forcing (Fig.~\ref{xi}a, the predicted band gap in $f\in [0.9,1.3]$~Hz is well observed experimentally and numerically, as the localization length $\xi$ decreases significantly within the band gap. Within this band gap, the localization effects are the strongest and the localization length is minimum ($\xi \sim 1$ m). On either side of the band gap, the experimental and numerical values of $\xi$ agree but are not well described by the linear theory as weakly nonlinear effects probably occur even with $\epsilon=0.015$. It is a consequence of the finite amplitude of the waves, as confirmed by simulations (solid colored lines). For the random bathymetry (Fig.~\ref{xi}b), the experimental and numerical values of $\xi$ show a larger forbidden band because of the presence of multiple reflections and are well described by the linear theory (solid line) in the case of a weak forcing. 

For a strong forcing (nonlinear waves with $\epsilon=0.040$), in both the periodic and random cases, a strong decrease of $\xi$ appears regardless of $f$, meaning that localization effects are stronger for nonlinear waves. This observation is stressed in Fig.~\ref{xi}a and Fig.~\ref{xi}b where experimental and numerical results are similar and in Fig.~\ref{xi}c and Fig.~\ref{xi}d, where the experimental localization length $\xi$ is shown as a function of nonlinearity, i.e., wave steepness $\epsilon$, for three different frequencies. For a frequency inside the theoretical band gap (e.g., $f=1.1$~Hz), $\epsilon$ has almost no effect on $\xi$ as waves are already strongly localized in the linear case. For frequencies outside the band gap (e.g., $f=0.7$ or 1.5~Hz), the localization length $\xi$ decreases significantly with increasing nonlinearity. This thus shows, for the first time experimentally, that nonlinearity enhances the localization phenomenon of hydrodynamics surface waves. Such nonlinear effects have been predicted theoretically for other systems~\cite{Frohlich1986,Albanese1988}.

\begin{figure}[t!]
    \centering
    \includegraphics[width=1\linewidth]{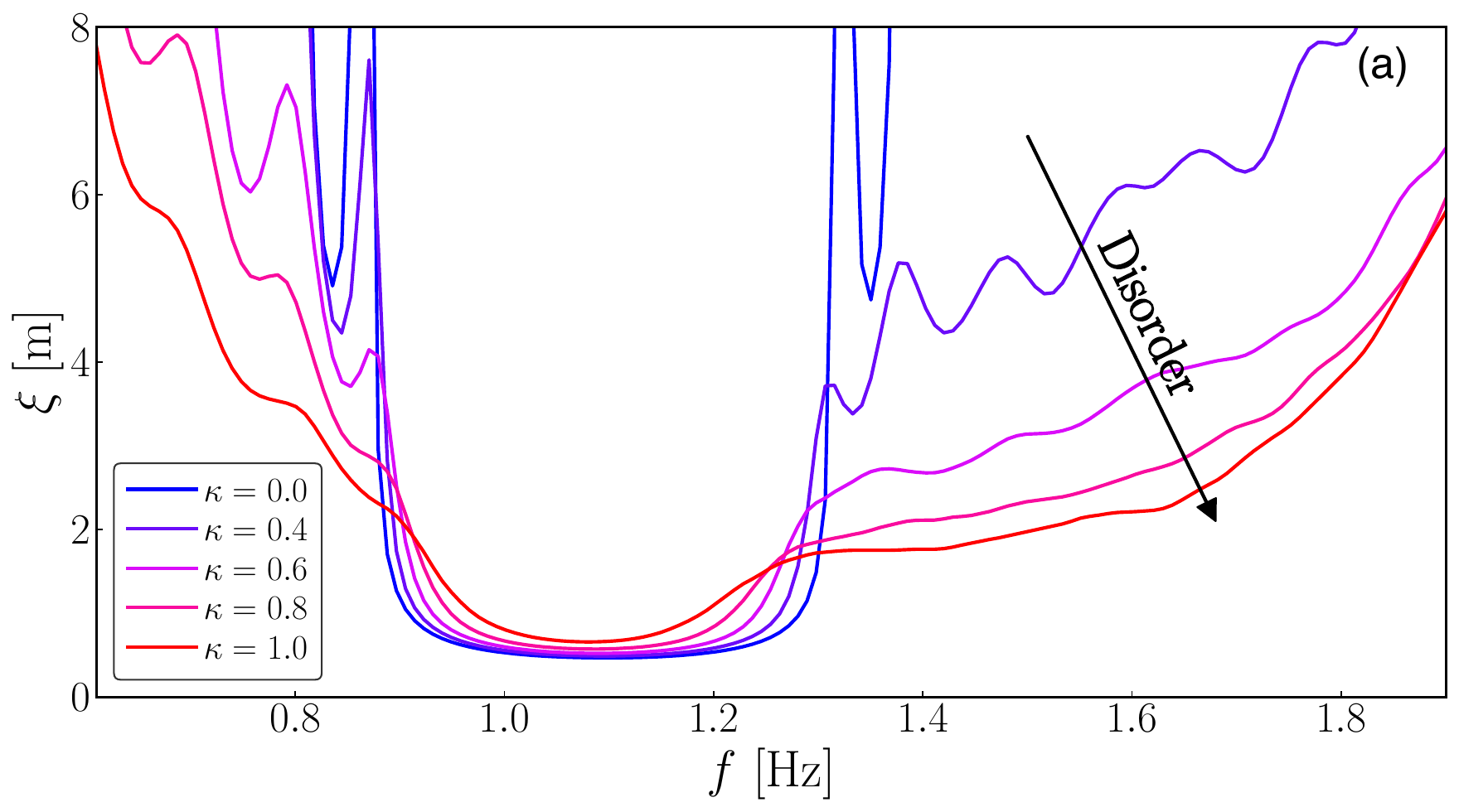}\\%
    \includegraphics[width=1\linewidth]{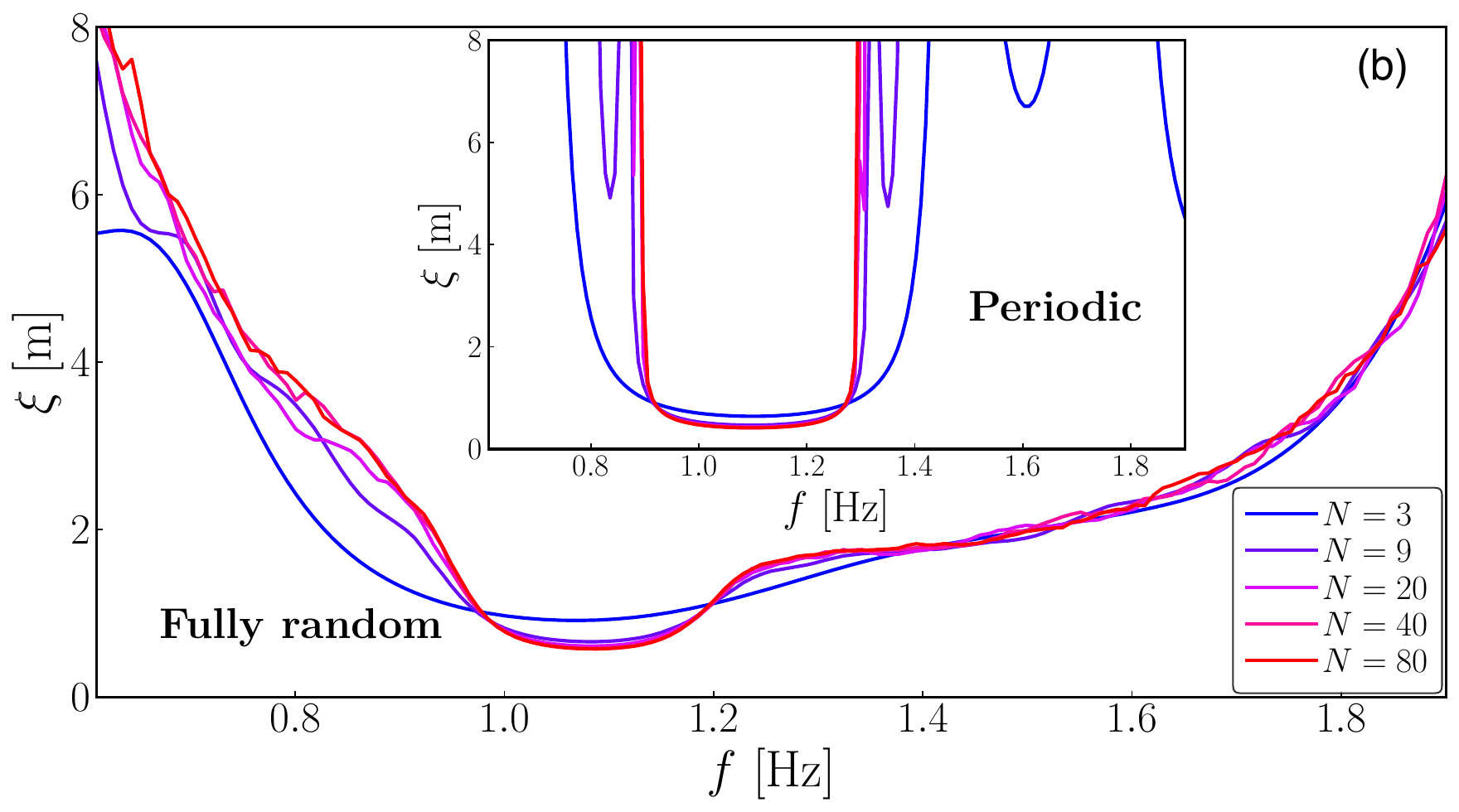}
    \caption{\textbf{Effects of disorder level and system finite size.} \textbf{(a)} Effect of the disorder level $\kappa$ on the localization length $\xi$ versus frequency $f$. $\kappa$ is increased (arrow) from 0 (periodic case) to 1 (completely randomness).  Theoretical curves for $N=9$ and $L=25$ cm (values of the experimental parameters). Blue ($\kappa=0$) and red ($\kappa=1$) curves correspond to the theoretical ones in Fig.~\ref{xi}a and~\ref{xi}b, respectively. \textbf{(b)} System finite-size effects on the localization length $\xi$ when the number $N$ of bars, and thus the system size, $L_x$, increases (solid lines from blue to red) for the periodic case (inset, $\kappa=0$) and the complete random case (main, $\kappa=1$). Theoretical curves for $L=25$ cm.}
    \label{Disorderlevel}
\end{figure}

We now discuss the relevant lengths involved in the problem. Typically, one has $L\lesssim \lambda \lesssim \xi < L_x \ll l_d$ as expected for waves to be localized, with $L=0.25$ m the disorder scale, $\lambda$ the wavelength, $\xi$ the localization length, $L_x=4$ m the system size, $l_d\sim 20$ m the dissipative length. These conditions are well fulfilled for most probed frequencies (see Supp.~Sect.~7).  Note that, for the periodic case, one has $\lambda \gtrsim \xi$ in the band gap, which explains the fact that the experimental values of $\xi$ are slightly higher than the predicted ones in the gap (see Fig.~\ref{xi}a).  Note also that to measure $\xi$, its value has to be less, or at least not too large, compared to $L_x$, which is valid in our experimental frequency range. The choices of the experimental parameters ($a$, $d$, $L$, $L_x$, $h$, $h_1$) come from a compromise made to respect the above constraints.

\vspace{0.2cm}
\noindent \textsf{Effects of disorder level and system finite size}\\ 
We now study the role of the level of disorder $\kappa$ on the localization length $\xi$ using the linear theory computed from the transfer matrix method of Eq.~\eqref{loca_th_eq}. When $\kappa$ is increased from 0 (periodic case) to 1 (completely randomness), Fig.~\ref{Disorderlevel}a shows that the band gap corresponding to Bragg scattering of a periodic lattice (blue curve) is gradually lost when the randomness is strong enough ($\kappa\gtrsim 0.6$). Disorder has a strong effect, resulting in a significant widening of the $\xi(f)$ curve, as all eigenstates are localized and not simply those within the band gap. This thus shows that the above presented experimental results with a strong random bottom ($\kappa=1$) are well ascribed to disorder-driven Anderson localization. Only the minimum of the $\xi(f)$ curve is reminiscent of the band gap. We thus have evidenced a smooth transition between Bragg scattering ($\kappa=0$) and Anderson localization ($\kappa=1$) when the disorder level increases, as also reported for 1D acoustics waves~\cite{Depollier1986,Desideri1993}.

Finally, we theoretically study the effects of the finite size of the system by changing the number $N$ of bars, and thus the total length $L_x$ of the system, to keep a constant distribution ($L+\delta$) of distances between bars, and thus a constant band-gap frequency. Finite-size effects, generating oscillations in the $\xi(f)$ curve, are predicted to be almost totally lost when $N\gtrsim10$ as shown in Fig.~\ref{Disorderlevel}b for the periodic case (inset, $\kappa=0$) and completely random case (main, $\kappa=1$). We can therefore conclude that in both cases the experiments performed here with $N=9$ are only slightly affected by finite size effects.

\vspace{0.2cm}
\noindent \textbf{Discussion}\\
Anderson localization has been observed experimentally with cold atoms~\cite{Billy2008,Roati2008,KondovScience2011,JendrzejewskiNatPhys2012} and classical waves (light waves~\cite{WiersmaNature1997,StorzerPRL06,PertschPRL04,SchwartzNature2007,Lahini2008}, microwaves~\cite{DalichaouchNature91,ChabanovNature00}, acoustic waves~\cite{Hodges1982,Depollier1986,Hu2008})  but mainly for linear waves whereas experimental studies with nonlinear waves are scarce~\cite{Hopkins1996,Hopkins1998,Mckenna1992,PertschPRL04,SchwartzNature2007,Lahini2008}. Here, we have reported experimentally, for the first time, the role of the nonlinearity on wave localization of hydrodynamic surface waves. We have shown that nonlinearity significantly increases wave localization regardless of the wave frequency except inside the reminiscent band gap where the localization length keeps its linear value. We have also shown the experimental evolution of the localization length with nonlinearity, which has never been reported previously with any wave.  The localization length is found to decrease with the nonlinear parameter. To do so, we have used the most recent available, full space-and-time-resolved wavefield measurement. In contrast, most previous experimental studies focused on measurements of the reflection and/or transmission coefficient of linear waves before/after the disordered medium. This also provides a much more accurate characterization of the linear gravity-wave localization regime than that reported in the 80s~\cite{Belzons1987,Belzons1988}. To our knowledge, no theoretical works taking into account these nonlinear effects have been derived so far. We have also reported experimentally the first evidence of the macroscopic analog of Bloch's dispersion relation in the case of hydrodynamic surface waves over a periodic bottom. These results are supported by our numerical simulations of the fully nonlinear Boussinesq equations.  We also found theoretically a smooth transition between Bragg scattering (periodic case) and localization phenomenon (random case) when the level of the disorder increases. The system's finite-size effects on the localization length are also reported.  Finally, our experimental system, currently limited to cases of linear and nonlinear continuous waves, thus offers a convenient platform to tackle the question of Anderson localization of nonlinear structures (as nonlinearity can be controlled here), such as solitons~\cite{Li1988, Kivshar1990, Mei2004}, nonlinear pulses~\cite{Hopkins1996,Hopkins1998,PikovskyPRL08,IvanchenkoPRL11}, 
and the possible existence of solitons within the band gap~\cite{ChenPRL1987,MorschRMP2006}. To what extent our work could be linked to the occurrence of extreme waves induced by variable bathymetry is also of primary importance in oceanography~ \cite{Trulsen2012,Ducrozet2017,Michel2022}.\\

\vspace{0.4cm}
\noindent \textbf{Methods}\label{Methods}\\
\noindent \textsf{Experiments}\\
The experimental setup is described in the main part. A linear motor (LinMot PS01-37x120F) drives the wavemaker within the glass canal. The five cameras are Basler ace U (acA3088-57um, 3088$\times$2064 pixels$^2$) equipped with 8 mm lenses (Basler C23-0824-5M-P), positioned 90 cm from the canal. The vertical and horizontal resolutions are 0.28 mm/pixel. The cameras deliver up to 80 frames per second (20 frames per second are used here). Image acquisition, processing, and analysis are carried out using Matlab software. \\

\noindent \textsf{Numerical methods}\\
Numerical simulations are performed using the fully-nonlinear Boussinesq equations solver, as part of Basilisk software~\cite{Popinet2015}. The rectangular bars are implemented as trapezoids, with a bottom base of $10$ cm, and a top base of $6$ cm, which we consider to give an ``effective" width of 8 cm. The size of the domain and all other dimensions are the same as in the experiment. A high resolution is chosen, with 1024 points. We impose a radiative boundary condition on one side, with its amplitude chosen carefully to match the nonlinear parameter, $\epsilon$, of the experiments. All of the data analysis procedures match the experimental ones.\\

\noindent \textbf{Data availability}\\
Data supporting key conclusions of this work are included within the article and Supplementary information. All raw data that support the findings of this study are available as a Source Data file. Source data are provided with this paper.\\


\vspace{0.2cm}
\noindent \textbf{Acknowledgments}\\
We thank J. Moukhtar for the technical discussions, and A. Di Palma and Y. Le Goas, for technical help. We thank C. Mora for suggestions. This work was supported by the Simons Foundation MPS No.~651463 (E.F.) Wave Turbulence (USA) and the French National Research Agency [ANR SoGood Project No.~ANR-21-CE30-0061-04 (E.F.) and ANR Lascaturb Project No.~ANR-23-CE30-0043-02 (E.F.)].\\

\noindent \textbf{Author informations}\\
\noindent \textsf{Contributions}\\
E.F. conceived the idea of the research. G.R., F.N., and E.F. designed the experiments. G.R. performed the experiments and carried out the data analysis. F.N. developed the code and performed the numerical simulations and the theoretical calculations. E.F. directed the project and received the funding. All authors discussed the experimental, numerical, and theoretical results. G.R. wrote the first draft of the manuscript. All authors outlined the manuscript content, and reviewed and edited the manuscript.\\


\noindent \textbf{Ethics declarations}\\
\noindent \textsf{Competing interests}\\
The authors declare no competing interests.\\


\end{document}